\pdfoutput=1

\RequirePackage[hyphens]{url}
\documentclass[a4paper,11pt]{article}
\usepackage{jheppub}
\usepackage[section]{placeins}
\usepackage{bm}
\usepackage{appendix}
\usepackage[hyphens]{url}
\usepackage{hyperref}

\usepackage{float}
\usepackage{subfig}
\usepackage{cancel}
\usepackage{lscape}

\usepackage{color}
\usepackage[usenames,dvipsnames,svgnames,table]{xcolor}

\usepackage{amsmath}
\usepackage{amssymb}
\usepackage{graphicx}
\usepackage{slashed}
\usepackage{soul}
\usepackage{multirow}

\def\lsim{\mathrel{\raise.3ex\hbox{$<$\kern-.75em\lower1ex\hbox{$\sim$}}}}
\def\gsim{\mathrel{\raise.3ex\hbox{$>$\kern-.75em\lower1ex\hbox{$\sim$}}}}

\newcommand{\bea}{\begin{eqnarray}}
\newcommand{\eea}{\end{eqnarray}}

\newcommand{\be}{\begin{equation}}
\newcommand{\ee}{\end{equation}}

\title{Personal Ultraviolet Respiratory
 Germ Eliminating Machine~(PUR$\diamond$GEM) for COVID-19}
\author[a]{Nausheen~R.~Shah,}
\author[b]{Ismar~Masic,}
\author[c]{Chris~Jones,}
\author[d]{Ritesh~Gupta}

\affiliation[a]{Department of Physics \& Astronomy, Wayne State University, Detroit, MI 48201, USA}
\affiliation[b]{88090 Immenstaad, Germany}
\affiliation[c]{Vanabot, London SW18 1SS, UK}
\affiliation[d]{Houston, TX 7701, USA}

\emailAdd{nausheen.shah@wayne.edu}

\preprint{WSU-PHY-2001}

\abstract{
The current COVID-19 pandemic has highlighted the need for cheap reusable personal protective equipment. The disinfection properties of Ultraviolet~(UV) radiation in the 200-300 nm have been long known and documented. Many solutions using UV radiation, such as cavity disinfection and whole room decontamination between uses,  are in use in various industries, including healthcare. Here we propose a portable wearable device which can safely, efficiently and economically, continuously disinfect inhaled/exhaled air using UV radiation with possible 99.99\% virus elimination. We utilize UV radiation in the 260 nm range where no ozone is produced, and because of the self-contained UV chamber, there would be no UV exposure to the user. We have optimized the cavity design such that an amplification of 10-50 times the irradiated UV power may be obtained. This is crucial in ensuring enough UV dosage is delivered to the air flow during breathing. Further, due to the turbulent nature of airflow, a series of cavities is proposed to ensure efficient actual disinfection. The Personal Ultraviolet Respiratory Germ Eliminating Machine~(PUR$\diamond$GEM) can be worn by people or attached to devices such as ventilator exhausts/intakes, or  be used free-standing as a portable local air disinfection unit, offering modularity with multiple avenues of usage. Patent pending.   
}

\begin{document}

\maketitle

\section{Introduction} \label{sec:intro}

Current COVID-19 pandemic has highlighted the need for cheap reusable personal protective equipment. In particular, recent studies~\cite{van2020aerosol, fears2020comparative} showing the stability~($\gtrsim$ 16 hours) of aerosolized SARS-COV-2 make air disinfection one of the highest current priorities.  

The disinfection properties of Ultraviolet~(UV) radiation in the 200-300 nm have been long known and documented. Many solutions using UV radiation, such as cavity disinfection and whole room decontamination between uses,  are in use in various industries, including healthcare. Here we propose a portable wearable device which can safely, efficiently and economically, continuously disinfect inhaled/exhaled air using UV radiation with possible 99.99\% virus elimination. We utilize UV radiation in the 260 nm range where no ozone is produced, and because of the self-contained UV chamber, there would be no UV exposure to the user. We have optimized the cavity design such that an amplification of 10-50 times the irradiated UV power may be obtained. This is crucial in ensuring enough UV dosage is delivered to the air flow during breathing.  Further, due to the turbulent nature of airflow, a series of cavities is proposed to ensure efficient actual disinfection. The Personal Ultraviolet Respiratory Germ Eliminating Machine~(PUR$\diamond$GEM) can be worn by people or attached to devices such as ventilator exhausts or  be used free-standing as a portable local air disinfection unit, offering multiple avenues of usage. 

The elimination of an organism due to UV radiation is highly dose dependent. There has been a great deal of investigation on the optimal dose required, showing a huge range of efficacious dosages administered~(See for e.g. Ref.~\cite{memarzadeh2010applications} and references within). However, investigating these studies, we found that there were sometimes confounding factors leading to inconsistent results. We found that the most pertinent investigations were carried out in Refs.~\cite{walker2007effect, mcdevitt2012aerosol, buonanno2017germicidal, welch2018far, newcorona}. Most recently, Ref.~\cite{newcorona} showed that 222 nm  UV radiation is able to inactivate various coronaviruses to the 3-log level with a dose of $\lesssim$ 2 mJ/cm$^2$. Previously, Ref.~\cite{welch2018far} showed that the H1N1 virus requires a dose of 4 mJ/cm$^2$ to give a 4-log (99.99\%) reduction using 222 nm radiation. In Ref.~\cite{mcdevitt2012aerosol} it was shown that irradiation by 254 nm UV provided similar elimination of H1N1. For the Murine Hepatitis Virus coronavirus, a similar dose requirement was found when irradiated with 254 nm UV~\cite{walker2007effect}. In previous work, the authors of  Ref.~\cite{newcorona,welch2018far} have shown that 254 nm and 222 nm radiation gives similar results for bacterial targets~\cite{buonanno2017germicidal}, even though the precise cause of elimination is different~\cite{beck2016comparison}. Very recently, a preliminary study shows that a 3-log reduction for SARS-COV2 may be obtained for similar values of 3.7 mJ/cm$^2$~\cite{Bianco2020.06.05.20123463}.  This result is inline with expectations due to the structural similarities in the various viruses. For the long term user,  one would optimally want enough UV radiation to also eliminate bacteria and fungal pathogens, which seems to require an order of magnitude larger dose for similar disinfection levels~\cite{memarzadeh2010applications}. However, due to the larger size of bacteria/fungi compared to viruses, bacteria/fungi tend to get filtered very efficiently with even low quality fabric masks. Hence, it may be more efficient to consider a reusable, washable filter in conjunction with UV disinfection to provide optimal protection from all pathogens.    

There are many UV sources available commercially in a wide range of both power and wavelength. While Mercury based cold cathode lamps~(CCL) are cheaper currently, LEDs have been dramatically reducing in price, and with the UN Minamata Convention on Mercury~\cite{UN}, it is expected that mercury lamps will get phased out soon. Additionally, LED lifetimes are between 10 - 20,000 hrs. Current LED pricing for 250-280 nm wavelength $\lesssim$ \$0.2/mW~\cite{mWplot, klaranprice}. Further, peak germicidal efficiency has long been considered to be 254 nm, but this is because 254 nm corresponds to the highest peak of the mercury spectrum. However, it is well known that the peak absorbance of RNA/DNA is at 260 nm. Hence, it stands to reason that higher elimination efficiency may be achieved via irradiation of viruses and bacteria closer to 260 nm, a wavelength that can be achieved if using LEDs rather than mercury based lamps.

The challenge in any device proposing to continuously disinfect air is to ensure the practical delivery of the required UV dosage given the rate of air flow dictated by a person breathing while being economically feasible. The average rate of air inhaled/exhaled under light to moderate exertion is $\sim$ 20 liters/minutes~\cite{OxRate, janssen2003principles, tipton2017human}, giving a volume flow rate of $V_F = 333 ~\rm{cm}^3$/s. The dose delivery of any such device is proportional to the power of the UV source, $P_{UV}$, and inversely proportional to $V_F$. In the PUR$\diamond$GEM we optimize the geometry of the UV cavity such that the dose delivered by a UV source of power $P_{UV}$ is significantly enhanced due to multiple reflections inside the chamber, and show that a nominal dose of 4 mJ/cm$^2$ for $V_F$ of air can be practically delivered with a 20 mW UV LED. Further, due to the turbulent nature of airflow, a series of cavities is proposed such that the actual efficient disinfection may be obtained. Details of disinfection obtained in series and prototype development will be presented in a companion publication~\cite{prototype}.  

\begin{figure}[thbp]
   \begin{centering}
       \includegraphics[width = 2.5in]{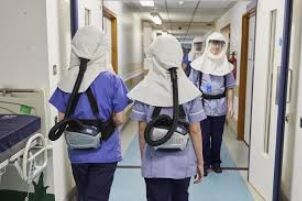}
      \hspace{.5in}
      \includegraphics[width = 2.5in]{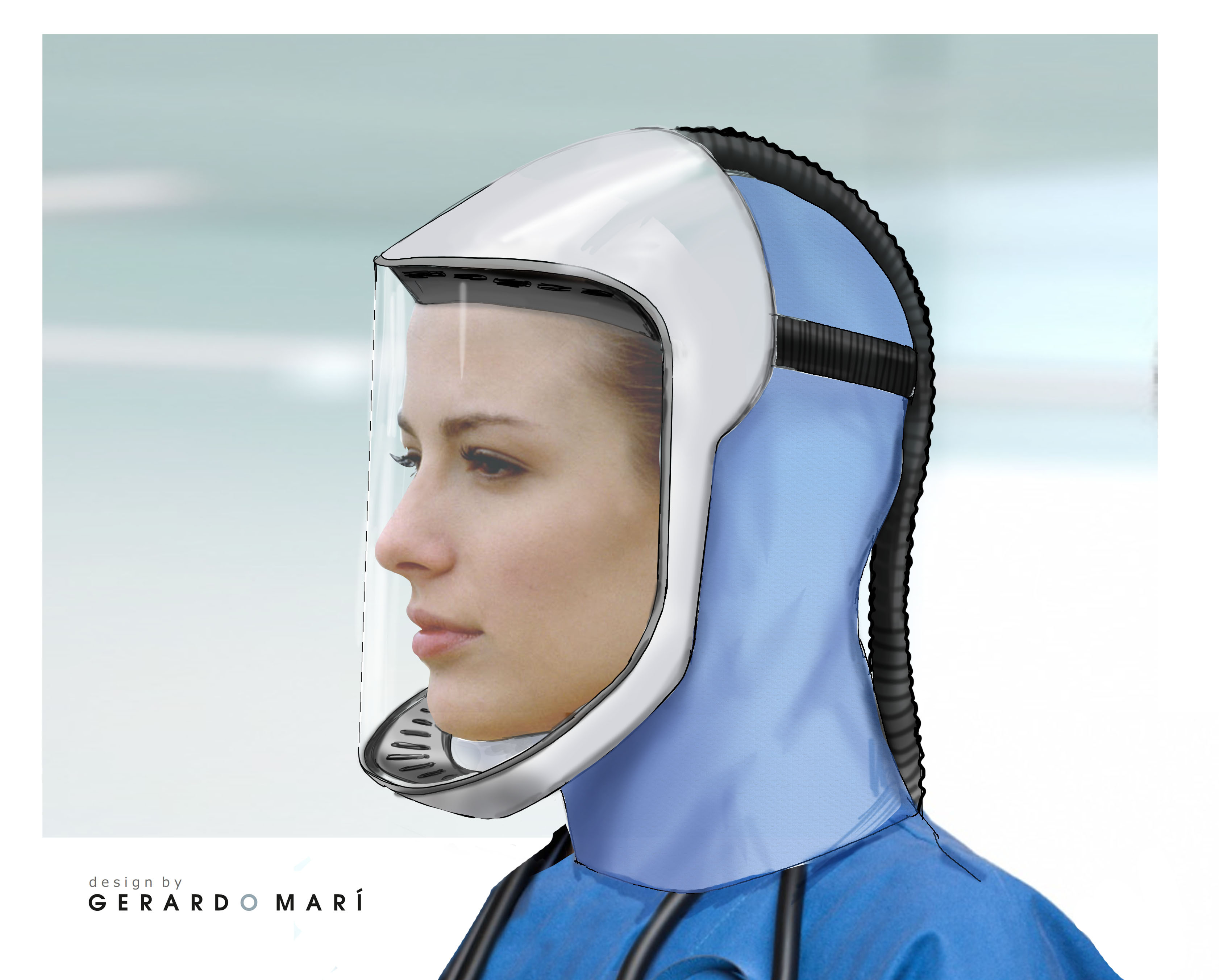}\\
(a) \qquad \qquad \qquad \qquad \qquad \qquad \qquad \qquad \qquad \qquad (b)
    
      \caption{(a) Respirator hoods attached to HEPA filters used by hospital workers in the UK~\cite{HEPAhoodPics}. (b) Visualization of a possible helmet that could be attached to the PUR$\diamond$GEM~\cite{OurHelm}.}
      \label{fig:helmet}
   \end{centering}
\end{figure}

We consider that the largest size contraption that could be practically carried/strapped to a person for long periods of time not exceed 20 cm $\times$ 20 cm $\times$ 40 cm, and must be lightweight. We envision that the PUR$\diamond$GEM will be hooked up to either a hood or a helmet, examples of these are shown in Fig.~\ref{fig:helmet}, but could be connected to a sealed mask if desired. A reasonable and cheap air seal may be obtained for example by using fashion tape or a velcro strap on the head piece appropriately. There will be low power fans blowing air to and from the UV chamber, so there doesn't need to be a completely air tight seal around the face as long as positive pressure is maintained.~\footnote{If used with sealed mask, fan assembly may not be necessarily required, however, the comfort level of the user would be reduced. Additionally, some sort of ventilation system may be needed to prevent CO$_2$ buildup~\cite{birgersson2015reduction}.} This makes this system continuously usable for long periods of time comfortably. All exposed surfaces of the PUR$\diamond$GEM system should be smooth, so can be easily cleaned by wiping with standard disinfecting solutions.      

In this article we propose a schematic for the PUR$\diamond$GEM UV chamber optimizing for practical considerations listed above. The PUR$\diamond$GEM can be configured as 1-way or 2-way air flow, corresponding to disinfecting only one way, or both inhalation and exhalation. For the same power delivered, this offers a trade-off between UV dose versus in and out disinfection. The next section details the PUR$\diamond$GEM UV chamber design which can be simply manufactured. We reserve Sec.~\ref{sec:conc} for discussion, summarizing our designs and next steps. The Appendix lists various possible sources for materials required.

\section{The PUR$\diamond$GEM} \label{sec:mod}

The crucial observation we make and utilize in our UV disinfection chamber design is the fact that effectively infinite reflections of light inside the chamber significantly enhance the initial UV source power irradiating air flow through the cavity. This is achieved practically by ensuring that the interior surface of the cavity is spherical and made of high reflectance material.

We consider the {\it minimal} possible average dose that could be received by air in the time duration of the air flow through the disinfection cavity, including an enhancement factor $E_{cav}$ due to reflections:
\be \label{eq:minD}
D_{min}= \frac{E_{cav}P_{UV}}{A_{irrad}}\frac{V_{cav}}{V_F}\;,
\ee
where $A_{irrad}$ is the maximum surface area that $P_{UV}$ irradiates, and $V_{cav}$ is the volume of the cavity. Clearly to maximize $D_{min}$, one wants to minimize $A_{irrad}$ for some given $V_{cav}$. 

For an isotropic diffuse source, the initial power irradiating a given surface in a cavity will generically get suppressed by $1/d^2$, where $d$ is the distance from the source. A sphere minimizes the surface area given a volume, and is the geometry used in our model. We note here that a cylinder of radius $r$ and height $2r$ (or vice versa) and a cube of side $2r$ would have the same $V_{cav}/A_{irrad}$ as a sphere of radius $r$. However, the radiation density in a cylinder or cube compared to a sphere is not expected to be as uniform for a point source such as a UV LED, but may be comparable for a line source such as that provided by a CCL. Detailed computer simulations would need to be performed to verify precise dosage in a non-spherical cavity.

We envision the air flow ports in the UV chamber to be small compared to the surface area of the cavity. Hence we expect that the irradiation of air as it flows through the spherical cavity should be mostly uniform given $V_F$ and the size of the UV chamber. 

Finally, average 
travel time for a given air volume does not translate into actual UV power dose received leading to pathogen elimination directly. Instead one needs to compute the expectation value for disinfection using a weighted probability distribution for travel times through the cavity. Details will be provided in companion publication on prototype development~\cite{prototype}.

\subsection{The  Model} \label{sec:diff}
 
 \begin{figure}[hptb!]
   \begin{centering}
       \includegraphics[width = \textwidth, trim = 0cm 1cm 0.5cm 2.cm, clip]{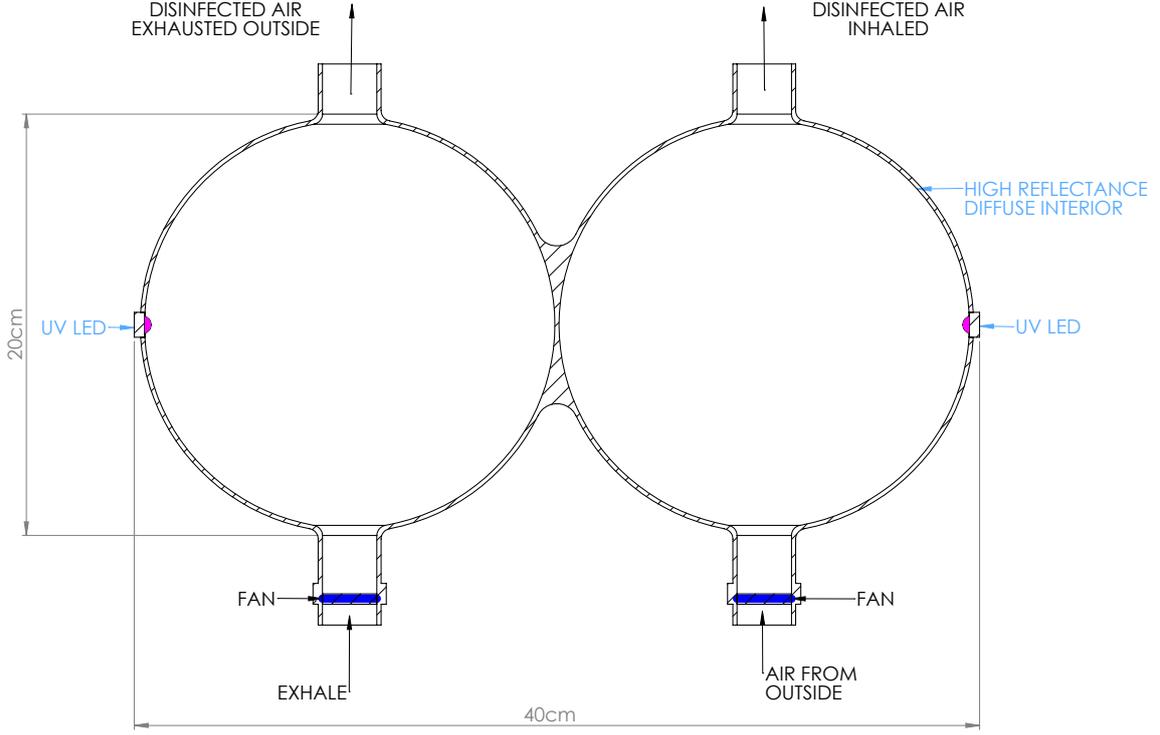}
      \caption{A schematic for the diffuse model configuration for the PUR$\diamond$GEM. A UV source (LED) irradiates the internal cavity uniformly due to high reflectance interior walls. Air flow at required rate is propelled by example placement of fans as shown. Drawings not to scale. See text for details. }
      \label{fig:diff}
   \end{centering}
\end{figure}

 A schematic of the  PUR$\diamond$GEM  is shown in Fig.~\ref{fig:diff}. A spherical cavity, made of high reflectance (diffuse) material is the main UV chamber. As an example material, Polytetrafluoroethylene (PTFE)~\cite{PTFE}  is a widely and cheaply available UV resistant material with $> 95\%$ reflectance~\cite{weidner1985laboratory}. Due to the high reflectance and spherical shape, the irradiance is expected to be uniform throughout the sphere.  A UV LED is mounted on the side, irradiating the entire sphere. There are several air inlets and outlets clustered at the two poles, configured to be furthest  away from each other to optimize diffuse air flow through the chamber.  In case of using loose fitting hoods, it may be also be desirable to have multiple fan speeds, such that the air flow rate may be changed depending on the exertion level of the user.  Of course this would reduce the dose for the same power source, since the minimal dose administered, Eq.~\ref{eq:minD}, is inversely proportional to $V_F$.  The input fan assembly may include a tight-weave cloth filter covering the opening to the chamber, acting primarily as a particulate filter~\cite{konda2020aerosol}. More sophisticated particulate filters may be used if desired. This fan assembly would be removable, and the cloth filter can be simply washed in  soap and water as needed. PTFE is highly dust repellent, so there is unlikely to be dust build up inside the chamber, however, there could be a UV sensor in the chamber which would alert the user if there is reduced UV dosage in case of dust buildup or LED failure. Alternatively, a current sensor could be incorporated into the circuit design which alerts the user to current flow below a certain level, indicating LED/fan failure or low battery. The UV chamber can be opened, so that in case of dust buildup, it can be cleaned out for example with ethanol. For safety purposes, there would be sensors such that the device will be inoperable when either the UV cavity or air flow vents are exposed. It would be optimal if the air ports could be sealed off when the device is not in use to minimize dust accumulation within UV chamber. Finally, all components  should be enclosed within a smooth cavity.

Due to the reflectance of the walls, one gets a geometric enhancement factor for the UV power. This is reduced by any defects in the reflecting surface, such as those due to the surface area occupied by the LED and the air vents. Therefore, the effective reflectance $R_{eff}$ of the cavity is given by:
 \be
 R_{eff}= R\left(\frac{SA_{s}-A_{def}}{SA_{s}} \right)\; ,
 \ee  
 where $R$ is the reflectance of the walls, $SA_{s}=4\pi r^2$ is the surface area of the sphere of radius $r$,  and $A_{def}$ is the total area of defects. Practically one should limit $A_{def}$ to be less than $\sim5\%$ of $SA_{s}$. The enhancement due to effectively infinite reflections of light inside the cavity  gives a multiplicative factor to $P_{UV}$:
 \be
 E_{s}=\frac{1}{1-R_{eff}}\;.
 \ee
 
 The nominal dose $D_s$ is then computed as follows:
 \be
 D_s = \frac{E_{s} ~P_{UV}}{SA_s} \frac{V_s}{V_F} = \frac{P_{UV}}{3 V_F ({1-R_{eff}})} ~r \; ,
 \ee
 where $V_s=\frac{4}{3} \pi r^3$ is the volume of the sphere. We see that $D_s$ is proportional to the $r$. Considering our benchmark numbers of 
 \be
 V_{F} = 333~{\rm cm}^3/{\rm s}, ~P_{UV} = 20~{\rm mW}, ~R_{eff} = 0.95~(E_s=20), ~r=10~{\rm cm},
 \ee
 \be
 D_s = 4~\rm{mJ/cm^2}.
 \ee

The spherical shape of the diffuse model provides excellent spatial integration. Since there are minimal radiation hot or cold spots, the UV dose received by in-passing air volume becomes effectively independent of path taken through the sphere. However, travel time distributions, not average time, must be taken into account to accurately predict actual disinfection, and in general will reduce the actual pathogen elimination significantly. We note that having a series of spheres will effectively integrate the paths temporally, allowing for the achievement of much higher levels of actual disinfection. Taking into account the travel time distribution function, a series of $n$ spheres with the same total linear dimensions~(radius $r/n$) as compared to a single sphere~(radius $r$) and same total UV power source would significantly reduce footprint of the device as well as material costs of cavities. Details are be provided in companion publication on prototype development~\cite{prototype}.

\section{Discussion} \label{sec:conc}

The PUR$\diamond$GEM can provide safe, practical, economically feasible, continuous personal disinfection of air for $\sim$ 99.99\% of viral particles. Disinfection from UV has been shown to be efficacious for a wide range of wavelengths for a whole host of organisms. The crucial and novel part of our design is the possible 10-50 times amplification of the initial UV source power in a cavity. Further, a series of cavities will provide sufficient dosage to be delivered to air flow while breathing, allowing effective actual pathogen elimination. LEDs in the 265~nm wavelength may be optimal economically due to ease of availability. Currently UV LEDs of up to $\sim$100~mW are commercially available for such wavelengths. Mercury based CCLs are significantly cheaper, being easily available for powers $\sim$1 W for the same price point.

The UV cavities proposed here may be manufactured using several different ideas. We first note that high precision cavity tolerances are not required such as would be needed for precision integrating spheres used for optical power measurements. As long as the cavities are not severely deformed, small surface imperfections should approximately average out and the UV dose delivered to flowing air should be approximately uniform.  Aluminum cavities can be easily made at low costs but would only give factors of $\sim 3$ enhancement in power. Our goal here is to get enhancement factors of $\gtrsim \mathcal{O}(10)$ in an economically feasible way. As mentioned before, PTFE has a reflectivity $\gtrsim 95\%$ at the wavelengths we are considering, corresponding to an enhancement factor of $\gtrsim 20$. PTFE is a depth reflector rather than a surface reflector. However, only $\sim$ 0.024 g/cm$^2$ or a 0.1 mm deep layer of virgin PTFE is required for achieving optimal reflectivity~\cite{janecek2012reflectivity, Barabash:2017sxf}. Thin tapes and sheets of PTFE  are easily and cheaply available commercially. PTFE membranes may also be used~\cite{Porex}. These have a higher reflectance, but require a thicker layer, $\sim$ 2 mm, to achieve optimal reflectance.  These can be cut to size and then layered inside any cavity which is either cheaply  available commercially, or manufactured via 3D printing. Alternatively, the cavities may be coated with other high UV reflectance coatings such as compounds containing Barium Sulphate, which may potentially allow for an even higer reflectance~\cite{schutt1971formulation, schutt1974highly,knighton2005mixture, janecek2012reflectivity}.  
All other materials~(fans,  UV sensors etc.) are easily and cheaply  available. Finally, considering the power consumption from the UV sources and fans, we estimate that this system could be powered using a standard rechargeable 10 Ah 5V power bank, providing between 5-8 hours of continuous usage depending on precise power ratings of components. We list some possible material resources in the Appendix.

We have presented a benchmark configuration with a computed dosage nominally corresponding to 4-log reduction of the H1N1 virus. However, a series of such cavities will be required to provide such actual elimination~\cite{prototype}. Recent work supports that dosages for SARS-COV2 may be similar to H1N1. The benchmark cavities were configured to be the maximal dimensions that we imagine could be practically worn, and doses computed using the minimal power required to give the needed virus elimination for minimal air flow rate $V_F$ we thought acceptable~\cite{janssen2003principles}. For the model proposed, the computed dosages scale linearly with the largest dimension of the cavity and the UV power. With similar sized cavities as we propose, a UV source of 10 times the power may be needed to give similar elimination of other pathogens (bacterial or fungal). Given the low cost of both LEDs and CCLs, such power requirements do not seem problematic. However, a reusable particulate filter may provide more efficient elimination of such organisms. For the configuration presented, 2-way airflow may be converted to 1-way if needed, providing a a higher level of disinfection. Finally, a series of cavities~(comparing same total linear dimension and power) may provide a significant enhancement in actual virus elimination considering travel time distributions through UV cavities, as well as reducing fingerprint and material cost of cavities~\cite{prototype}.

There are many UV disinfection solutions being used commercially, specially in the healthcare industry, such as decontamination cavities and whole room irradiation between uses. Recently it has been proposed that 222 nm radiation could be used for the later purpose without endangering the occupants~\cite{buonanno2017germicidal, welch2018far, newcorona}. However, such measures are only effective on the areas directly exposed; UV radiation is generally very easily absorbed by any obstructions.
While personal protective equipment has always been a need in the healthcare industry, the current situation is exacerbated by their conventionally disposable nature, dictated by the need for infection control and degradation of particulate filters over use.  We present a personal protection solution which is safe, reusable and requires minimal maintenance, while providing superior protection. We estimate the price-point for the PUR$\diamond$GEM system~(including minimal hood/helmet) produced commercially to be $\sim \mathcal{O}($\$100). Single-use N95 masks cost $\sim$\$1-5 each; while N95 masks can be decontaminated with UVC, they degrade rapidly~\cite{n95}, leading to a poorer fit and higher possibility of virus transmission. Hence the economical feasibility of this device is immediately obvious. The PUR$\diamond$GEM can be deployed in the healthcare industry as well as in the common population, allowing society to resume its functionality safely. This is of paramount importance right now, given that even with promising preliminary results for a vaccine for COVID-19~\cite{pfizerVac, ModernaVac}, efficient population vaccination is not expected for $\gtrsim$ 1 year.

\acknowledgments
Idea of reflective spherical cavity, all design optimizations, analytic calculations, development of series configurations, and prototype development undertaken by N.R.S, patent pending. Both I.M. and C.J. performed initial numerical simulations to validate analytical calculations.  While both I.M and N.R.S had individually been thinking about a personal device using UV radiation for continuously disinfecting air, this collaboration was initialized through the Helpful Engineering initiative~\cite{HE}. We thank the many volunteers for their thoughtful discussions and feedback. This project originally began as a modular part of a larger initiative organized by R.G. to build a whole respirator system including helmet.

\appendix

\section{Possible Material Resources}
\begin{itemize}
\item LEDs \\
\url{https://www.klaran.com/images/Products/CIS_Klaran_WD_DS_032020.pdf}\\

\item CCLs \\
\url{https://www.assets.signify.com/is/content/PhilipsLighting/fp927900504007-pss-global}\\

\item Heat Sink\\
\url{https://www.digikey.com/product-detail/en/wakefield-vette/19756-L-AB/345-1119-ND/3175833}\\

\url{https://www.alphanovatech.com/en/c_lpd50e.html}

\item Premanufactured Cavities\\
\url{https://www.webstaurantstore.com/fat-daddios-pha-65-proseries-6-1-2-x-3-1-4-hemisphere-anodized-aluminum-cake-pan/627PHA65.html}\\

\url{https://www.steelsupplylp.com/sku/102376}\\

\url{https://www.amazon.com/Acrylic-Dome-Plastic-Hemisphere-Pre-Drilled/dp/B01LZMZG5M/ref=sr_1_1?dchild=1&keywords=supreme\%2Btech\%2Bhemisphere&qid=1589045455&s=home-garden&sr=1-1-catcorr&th=1}\\

\url{https://www.amazon.com/gp/product/B016V20IJS/ref=ppx_yo_dt_b_asin_title_o06_s00?ie=UTF8&psc=1}\\

\item PTFE Membranes\\
\url{https://scientificfilters.com/roll-stock-membranes/ptfe_}\\

\item PTFE Sheets and Films\\
\url{https://techneticsptfe.com/products/ptfe-tapes-films/relic-wrap-archival-packaging-storage}

\url{https://www.eplastics.com/PTFENAT0-010X12}\\

\url{https://www.pur-sheet.com/}\\

\url{https://www.amazon.com/Soles2dance-Industrial-Strength-self-Adhesive-Backing-TEFLON06x12-3M-S/dp/B01DZ7G0H6/ref=sr_1_22?dchild=1&keywords=3M\%2BPTFE\%2Bfilm\&qid=1589043147\&sr=8-22\&th=1\&psc=1}

\item PTFE Tape\\
\url{https://www.homedepot.com/p/1-2-in-x-520-in-Thread-Seal-Tape-31273/202206819}
%
%

\item PTFE Bonding agent\\
\url{https://www.amazon.com/Loctite-Plastics-Bonding-Activator-681925/dp/B000Y3LHXW/ref=sr_1_12?crid=2IID09G2ANJF1&dchild=1&keywords=loctite+plastics+bonding+system&qid=1589043409&sprefix=loctite+plastic\%2Caps\%2C152&sr=8-12}\\

%

\item High Reflectance Coating\\
\url{https://digitalcommons.usu.edu/cgi/viewcontent.cgi?article=1010&context=cpl_techniquesinstruments}\\

\item Barium Sulphate\\
\url{https://www.carolina.com/specialty-chemicals-b-c/barium-sulfate-laboratory-grade-500-g/846950.pr?question=barium+sulphate}\\

\item Aluminum Paint\\
\url{https://www.amazon.com/gp/product/B00AENFZG2/ref=ppx_yo_dt_b_asin_image_o08_s00?ie=UTF8&psc=1}\\

%

\item UV sensors\\
\url{http://www.geni-uv.com/UVC-LED-Sensor.php}\\

\item Fans\\
\url{https://www.digikey.com/product-detail/en/sunon-fans/MF50151VX-B00U-A99/259-1829-ND/7691033}

\url{https://www.digikey.com/product-detail/en/sanyo-denki-america-inc/109BC12GC7-1/1688-1035-ND/6191751}

%


\end{itemize}

\bibliographystyle{JHEP.bst}
\bibliography{theBib}

\end{document}